\documentclass[10pt,conference,compsocconf,letterpaper]{IEEEtran}
\newtheorem{theorem}{Theorem}

\newtheorem{corollary}[theorem]{Corollary}

\newtheorem{lemma}[theorem]{Lemma}

\newtheorem{definition}{Definition}

\usepackage{cite}
\usepackage{graphicx}
\usepackage{subfigure}
\graphicspath{{../pdf/}{../jpeg/}{../eps/}}

\DeclareGraphicsExtensions{.pdf,.jpeg,.png}
\usepackage{amsmath}
\usepackage[ruled,vlined]{algorithm2e}
\usepackage{url}
\begin{document}
\title{Core Maintenance in Dynamic Graphs: A Parallel Approach based on Matching}

\author{\IEEEauthorblockN{
Na Wang, Dongxiao Yu, Hai Jin, Qiang-Sheng Hua, Xuanhua Shi, Xia Xie\\
\IEEEauthorblockA{Services Computing Technology and System Lab\\Big Data Technology and System Lab\\Clusters and Grid Computing Lab\\School of Computer\ Science and Technology.\\
Huazhong University of Science and Technology, Wuhan, 430074,\ China\\
Email: \{Ice$\_$lemon,dxyu,hjin,qshua,xhshi,shelicy\}@hust.edu.cn}
}}

\maketitle

\begin{abstract}
The core number of a vertex is a basic index depicting cohesiveness of a graph, and has been widely used in large-scale graph analytics. In this paper, we study the update of core numbers of vertices in dynamic graphs with edge insertions/deletions, which is known as the core maintenance problem. Different from previous approaches that just focus on the case of single-edge insertion/deletion and sequentially handle the edges when multiple edges are inserted/deleted, we investigate the parallelism in the core maintenance procedure. Specifically, we show that if the inserted/deleted edges constitute a matching, the core number update with respect to each inserted/deleted edge can be handled in parallel. Based on this key observation, we propose parallel algorithms for core maintenance in both cases of edge insertions and deletions. We conduct extensive experiments to evaluate the efficiency, stability, parallelism and scalability of our algorithms on different types of real-world and synthetic graphs. Comparing with sequential approaches, our algorithms can improve the core maintenance efficiency significantly.
\end{abstract}

\IEEEpeerreviewmaketitle

\section{Introduction}
As a basic index describing the cohesiveness of a graph, the core number of vertex has been broadly utilized in graph analytics. Specifically, in a graph $G$, the $k$-core is the connected subgraph in $G$, such that each vertex in the subgraph has at least $k$ neighbors. The core number of a vertex $v$ is then defined as the largest $k$ such that there exists a $k$-core containing $v$. The parameter of core number is also extensively used in a large number of other applications, to analyze the structure of a network, such as analyzing the topological structure of Internet \cite{app1}, identifying influential spreader in complex networks \cite{app3}, analyzing the structure of large-scale software systems \cite{app5}\cite{app8}, predicting the function of biology network \cite{app9}, and visualizing large networks \cite{app11} and so on. 

In static graphs, the computation of the core number of each vertex is known as the $k$-core decomposition problem, which has been extensively studied. The state-of-the-art algorithm is the one proposed in \cite{O(m)}. It can compute the core number of each vertex in $O(m)$ time and $m$ is the number of edges in the graph. However, in many real-world applications, graphs are changing continuously, due to edge/vertex insertions/deletions. In such dynamic graphs, many applications need to maintain the core number for each vertex in real-time. Hence, it is very necessary to study the core maintenance problem, i.e., update the core numbers of vertices in dynamic graphs.

\begin{figure}[!t]
 \subfigure[The core number change of $v_7$: from 1 to 2]{
    \label{cmp1_1} 
    \includegraphics[width=1.5in]{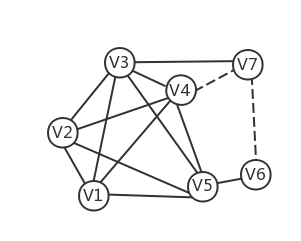}
    }
\subfigure[The core number change of $v_7$: from 1 to 3]{
    \label{cmp1_2} 
    \includegraphics[width=1.5in]{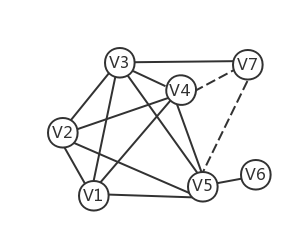}
    }
 \caption{In (a) and (b), two edges are inserted to vertex $v_7$ respectively, and the core number changes of $v_7$ in these two cases are 1 and 2 respectively.}
\label{insert2edges}
\vspace{-0.2in}
\end{figure}

\begin{figure} 
    \includegraphics[width=3.0in]{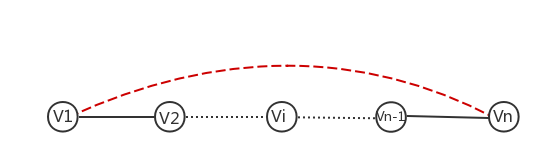}
    \vspace{-0.15in}
    \caption{After the insertion of edge $<$$v_1,v_n$$>$, all vertices increase the core number by one.}
\label{insert1edge}
\vspace{-0.2in}
\end{figure}

An intuitive way to solve the core maintenance problem is recomputing the core numbers of vertices after every change of the graph. But clearly, this manner is too expensive in large-scale graphs where there might be billions of vertices and trillions of edges. Another manner is just to find the set of vertices whose core numbers will be definitely changed and then update the core numbers of these vertices. However, this manner faces several challenges. First, 
the exact change value of core number of a vertex is not easy to determine, even if the same number of edges are inserted to a vertex, as shown in Fig. \ref{insert2edges}. Second, the set of vertices that will change the core number after a graph change is also hard to identify. As shown in Fig~\ref{insert1edge}, the core numbers of all vertices may change even if we only insert an edge to the graph.  

\begin{figure}
\centering
\includegraphics[width=2.9in]{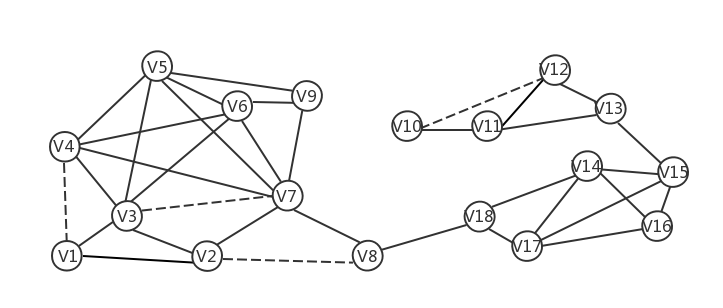}
\vspace{-0.15in}
\caption{Assume $<$$v_1,v_4$$>$,$<$$v_2,v_8$$>$,$<$$v_3,v_7$$>,<$$v_{10},v_{12}$$>$ will be inserted into the graph. \textbf{TRAVERSAL} algorithm in \cite{Sar2016Incremental} handle the inserted edges one by one. First for edge $<$$v_1,v_4$$>$, it will visit vertices $v_1,v_2$ and update core numbers for $v_1,v_2$ from 2 to 3. Then inserting $<$$v_3,v_7$$>$, it will visit $v_1,v_2,v_3,v_4,v_5,v_6,v_7,v_8,v_9$, and update core numbers of $v_3,v_4,v_5,v_6,v_7$ from 3 to 4. The same goes for $<$$v_2,v_8$$>$ and $<$$v_{10},v_{11}$$>$. During the process, $v_1$ and $v_2$ will be visited for multiple times, but their core numbers are only changed for once. However in our parallel algorithm, all four edges can be processed in parallel using three processors, and duplicate visitings of $v_1,v_2$ can be avoided.
}
\label{insertmatching}
\vspace{-0.2in}
\end{figure}

Due to the great challenge posed in solving the core maintenance problem in dynamic graphs, previous works all focus on the case that only one edge is inserted into/deleted from the graph. In this scenario, it is easy to check that the core number of each vertex can be changed by at most 1. Hence, the first challenge discussed above is avoided, and it only needs to overcome the second difficulty. When multiple edges are inserted/deleted, these edges are processed sequentially, and the core numbers of vertices are updated after each execution of the single-edge update algorithm. However, the sequential processing approach incurs extra overheads when multiple edges are inserted/deleted, since it may unnecessarily repeatedly visit a vertex, as shown in Fig. \ref{insertmatching}. And on the other hand, it does not make full use of the multi-core machine and distributed systems. Therefore, one natural question is whether we can investigate the parallelism in the edge processing procedure and devise parallel algorithm that suits to implement in multi-core machine and distributed systems. In this paper, we answer this question affirmatively by proposing parallel algorithms for core maintenance.   

The core maintenance in the scenarios of vertex insertion/deletion can be solved using algorithms for edge insertions/deletions. Take vertex insertion as an example. By setting the initial core number of the inserted vertex as 0 and executing an algorithm for edge insertion to handle the inserted edges generated by the inserted vertices, the core maintenance with respect to vertex insertion can be solved. Therefore, in this paper, we focus on the scenarios that there are only edges inserted into/deleted from the graph. Specifically, the core maintenance problem with respect to edge insertions and deletions are called the \emph{incremental} and \emph{decremental} core maintenance respectively. 

Our parallel algorithms are inspired by the single-edge insertion/deletion algorithms, such as those in \cite{Li2014TKDE}. To overcome the two difficulties discussed before, we first study the available set of edges whose insertions/deletions only make the core numbers of vertices change by at most one. If the feature of an available set can be determined, then when such an available set of edges are inserted/deleted, each edge can be processed in parallel: by finding the set of vertices whose core number changes due to each particular edge in the available set, the union of these sets of vertices are just those that will change core numbers after inserting the available set of edges, as inserting/deleting the available set can make each vertex change the core number by at most one. Based on this idea, we devise parallel algorithms consisting of two main steps: 1) split the inserted/deleted edges into multiple available sets, and 2) identify the vertices whose core numbers change after inserting/deleting each particular available set and update the core number of the vertices.  

Our contributions are summarized as follows.

\begin{itemize}
\item We show that if a matching (a set of edges any pair in which do not have common endpoints) is inserted/deleted, the core number of each vertex can change by at most one. 
\item Based on the structure of matching, we present parallel algorithms for incremental and decremental core maintenance respectively. Because a matching can contain an edge connected to each vertex that have edges inserted/deleted, the parallel algorithms reduce the iterations for processing all inserted/deleted edges from $m_c$ in sequential approaches to $\Delta_{c}+1$, where $m_c$ denotes the number of inserted/deleted edges and $\Delta_c$ denotes the maximum number of inserted/deleted edges connecting to a vertex. Though the number of inserted/deleted edges can be large, it is still very small in contrast with the number of vertices in a large-scale graph. Hence, $\Delta_c$ is small in real-world cases. Our algorithms will provide good parallelism in reality.
\item We then conduct extensive experiments on both real-world and synthetic graphs, to evaluate the efficiency, stability, parallelism and scalability of the proposed algorithms. The experiment results show that our algorithms exhibit good stability and scalability. Especially, our algorithms achieves better efficiency in handling graph changes of large size. Comparing with sequential algorithms, our algorithms speed up the core number update process on all datasets. In large-scale graphs, such as LiveJournal graph (refer to Table \ref{table_graph} in Section~\ref{sec:experiment}), the speedup ratio can be up to 3 orders of magnitude when handling the insertion/deletion of 20000 edges.
\end{itemize}

The rest of this paper is organized as follows. In Section~\ref{sec:relate}, we briefly review closely related works. In Section~\ref{sec:problem}, the problem definitions are given. Theoretical results supporting the algorithm design are presented in Section~\ref{sec:basis}. The incremental and decremental parallel algorithms are proposed in Section~\ref{sec:in} and Section~\ref{sec:de} respectively. In Section~\ref{sec:experiment}, the experiment results are illustrated and analyzed. The whole paper is concluded in Section~\ref{sec:conclusion}.

\section{Related Work}\label{sec:relate}
In static graphs, the core decomposition problem, which is to compute the core numbers of vertices, have been widely studied. In \cite{O(m)}, an $O(m)$ time algorithm was presented, where $m$ is the number of edges in the graph. This result is the state-of-the-art one. In \cite{massivenetwork}, an external-memory algorithm was proposed when the graph is too large to hold in memory. Core decomposition in the distributed setting was studied in \cite{TPDS13}. The above three algorithms were compared in \cite{SinglePC} under the GraphChi and WebGraph models. Parallel core decomposition was studied in \cite{ParK}.

In contrast, the work on core maintenance in dynamic graphs are fewer, and all known results focus on the single-edge insertion/deletion case. Specifically, in \cite{Sar2016Incremental}, it was shown that when one edge is inserted/deleted, the core number of any vertex can change by at most one. Based on this observation, a linear algorithm named \textbf{TRAVERSAL} was proposed to identify vertices that change core numbers due to the inserted/deleted edges. A similar result was also given in \cite{Li2014TKDE}. In \cite{I/OICDE16}, how to improve the I/O efficiency was studied, when computing and maintaining the core numbers of vertices. Distributed solutions for core maintenance of single-edge change were studied in \cite{AksuTKDE14} and \cite{DEBS16}.

\section{Problem Definitions}\label{sec:problem}
We consider an undirected, unweighted simple graph $G = (V,E)$, where $V$ is the set of vertices and $E$ is the set of edges. Let $n = |V|$ and $m = |E|$. For a vertex $u\in V$, the set of its neighbors in $G$ is denoted as $N(u)$, i.e., $N(u) = \{v\in V | (v,u) \in E\}$. The number of $u$'s neighbors in $G$ is called the degree of $u$, denoted as $d_G(u)$. So $d_G(u) =|N(u)|$. The maximum and minimum degree of nodes in $G$ is denoted as $\Delta(G)$ and $\delta(G)$ respectively.

We next give formal definitions for the \emph{core number} of a vertex and other related concepts.

\smallskip
\begin{definition}[\textbf{$k$-Core}]\label{de:kcore}
Given a graph $G=(V,E)$ and an integer $k$, the $k$-core is a connected subgraph $H$ of $G$, in which each vertex has at least $k$ neighbors, i.e., $\delta(H)\ge k$.
\end{definition}

\smallskip
\begin{definition}[\textbf{Core Number}]
Given a graph $G=(V,E)$, the core number of a vertex $u\in G$, denoted by $core_G(u)$, is the the largest $k$, such that there exists a $k$-core containing $u$. For simplicity, we use $core(u)$ to denote $core_G(u)$ when the context is clear.
\end{definition}

\smallskip
\begin{definition}[\textbf{Max-k-Core}]
The max-k-core associated with a vertex $u$, denoted by $H_u$, is the $k$-core with $k=core(u)$.
\end{definition}

In this work, we aim at maintaining the core numbers of vertices in dynamic graphs.
Specifically, we define two categories of graph changes: \emph{incremental}, where a set of edges $E'$ are inserted to the original graph, and \emph{decremental}, where a set of edges are deleted. Based on the above classification, we distinguish the core maintenance problem into two scenarios, as defined below.

\smallskip
\begin{definition}[\textbf{Incremental Core Maintenance}]
Given a graph $G=(V,E)$, the incremental core maintenance problem is to update the core numbers of vertices after an incremental change to $G$.
\end{definition}
\smallskip
\begin{definition}[\textbf{Decremental Core Maintenance}]
Given a graph $G=(V,E)$, the decremental core maintenance problem is to update the core numbers of vertices after a decremental change to $G$.
\end{definition}

The \emph{core number} of an edge is defined as the larger value of the core numbers of its endpoints. A set of edges $E'\subseteq E$ is called a \emph{matching}, if for each pair of edges in $E'$, they do not have common endpoints. If all edges in a matching have the same core number $k$, the matching is called a \emph{$k$-matching}.

We next give some notations that help identify the set of vertices which will change core numbers after graph change. Given a graph $G=(V,E)$, let $G'$ be the graph obtained after inserting/deleting an edge set $E'$ into/from $G$.

\begin{definition}[\textbf{Superior Degree}]\label{de:sd}
For a vertex $u\in V$, $v$ is a \emph{superior neighbor} of $u$ if $v$ is a neighbor of $u$ in $G'$ and $core_G(v)\ge core_G(u)$. The number of $u$'s superior neighbors is called the \emph{superior degree} of $u$, denoted as $SD_{G'}(u)$.
\end{definition}
 
\smallskip
The $SD$ value of a vertex $u$ represents the number of neighbors that have a core number no less than $core_G(u)$. It is easy to see that only superior neighbors of $u$ may affect its core number change.

\begin{definition}[\textbf{Constraint Superior Degree}]
The \emph{constraint superior degree} $CSD_{G'}(u)$ of a vertex $u$ is the number of $u$'s neighbors $w$ in $G'$ that satisfies $core_G(w) > core_G(u)$ or $core_G(w) = core_G(u) \land SD_{G'}(w) > core_G(u) $.
\end{definition}
\smallskip

The constraint superior degree of a vertex $u$ counts for two categories of neighbors that will affect the increase of $u$'s core number: the ones that have larger core numbers and those that have the same core number but have enough neighbors which may make themselves increase the core number. 

When the context is clear, we use $SD(u)$ and $CSD(u)$ to present the $SD$ and $CSD$ values of vertex $u$ in the current new graph.

In the insertion case, for a vertex $u$, if $CSD_{G'}(u) \le core_G(u)$, then $u$ cannot increase its core number, since it does not have enough support neighbors. We summarize this necessary condition formally as below and will use it to determine whether a vertex is impossible to increase the core number.

\begin{lemma}\label{le:inserne}
For a vertex $u$, if $CSD_{G'}(u) \le core_G(u)$, $v$ will not increase its core number.
\end{lemma}

Similarly, we can get a sufficient condition for a vertex to decrease the core number in the deletion case.

\begin{lemma}\label{corollary:sd}
For a vertex $u$ with core number $k$, if $SD_{G'}(u) < k$, then $u$ will decrease the core number.
\end{lemma}

\section{Theoretical Basis}\label{sec:basis}

In this section, we present some theoretical results that constitute the basis of our parallel algorithms. In particular, we first show that the core number of every vertex can change by at most one, if the inserted/deleted edges form a matching. And then we depict the feature of vertices whose core numbers will change after inserting/deleting a matching.

\subsection{Core Number Change of Vertices after Insertion/Deletion of A Matching}\label{Sec:matchinginsert}

The main results are given in Lemma \ref{Lem:matchinginsert} and Lemma \ref{Lem:matchingdelete} below. At first, we prove the following Lemma~\ref{Lem:decreaseby1}, which is useful in proving Lemma \ref{Lem:matchinginsert} and Lemma \ref{Lem:matchingdelete}.

\begin{lemma}\label{Lem:decreaseby1}
For a $k$-core $H=(V_H,E_H)$, $\delta(H)=k$, if after deleting a set of edges, $H$ becomes $H'$, where $H'$ is still connected, and the degree of each vertex in $H$ is decreased by at most 1, then $H'$ will be a $k$-$1$-core.
\end{lemma}

\begin{IEEEproof}
After deleting the edges, it can be concluded that $\delta(H')\geq k-1$. Then the result is obtained by Definition~\ref{de:kcore}.
\end{IEEEproof}

\begin{lemma}\label{Lem:matchinginsert}
Given a graph $G=(V,E)$, if a matching $E_M=\{e_1,e_2,...,e_d\}$ is inserted into $G$, then the core number of every vertex $w\in V$ can increase by at most 1.
\end{lemma}

\begin{IEEEproof}
First we assume that after the insertion, a vertex $w$ with core number $k$ increases the core number to $k+x$, where $x>1$. We denote the max-$(k+x)$-core of $w$ after the insertion is $H_w^+$ and the max-$k$-core of $w$ before insertion is $H_w$, then $\delta(H_w^+) = k+x$ and $\delta(H_w) = k$. It must be true that at least one of the inserted edges $e_i\in H^w_+$ for $1\le i\le d$, as otherwise $\delta (H_w)= k+x$ and $core_G(w)= k+x$, which is a contradiction. Let $Z=H_w^+\setminus E_M$, then $Z\subset G$. If $Z$ is connected, $Z$ will be a $k+x-1$-core by Lemma \ref{Lem:decreaseby1}, since the degree of every vertex in $H_w^+$ decreases by at most 1 due to the removal of a matching. This leads to a contradiction, since $k+x-1>k$ and $H_w$ is max-$k$-core in $G$. If $Z$ is disconnected, each connected component will be a $k+x-1$-core by Lemma \ref{Lem:decreaseby1}, since the degree of every vertex in each connected component decreases by at most 1 due to the removal of edges in $E_M$. This also leads to a contradiction. The result is then proved.
\end{IEEEproof}

Using a similar argument as that for proving Lemma \ref{Lem:matchinginsert}, we can get the result for deletion of a matching, as shown in the following Lemma \ref{Lem:matchingdelete}.

\begin{lemma}\label{Lem:matchingdelete}
Given a graph $G=(V,E)$, if a matching $E_M=\{e_1,e_2,...,e_d\}$ is deleted from $G$, then for every $w\in G$, $core(w)$ can decrease by at most 1.
\end{lemma}

\begin{figure}[!t]
\centering
\includegraphics[width=3.0in]{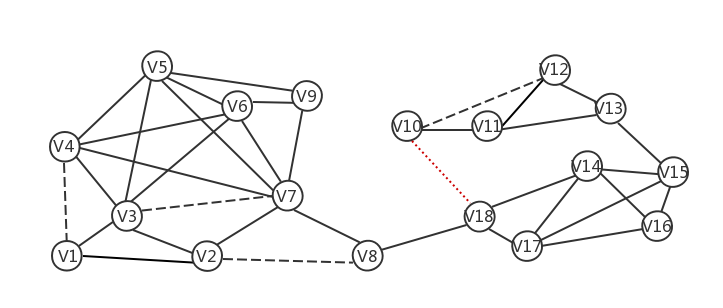}
\vspace{-0.1in}
\caption{A matching $\{<$ $v_1,v_4$ $>$,$<$ $v_3,v_7$ $>$,$<$ $v_2,v_8$ $>$,$<$ $v_{10},v_{12}$ $>\}$and an extra edge $<$ $v_{10},v_{18}$ $>$ are inserted. The core number of vertex $v_{10}$ will increase by 2, from 1 to 3.}
\label{morethanmatching}
\vspace{-0.2in}
\end{figure}

\textbf{Discussion.} In some sense, the matching may be the maximal structure of edges whose insertion/deletion only makes vertices change core number by at most 1, as shown by the example given in Fig. \ref{morethanmatching}. 

\begin{figure}[!t]
\centering
\includegraphics[width=2.5in]{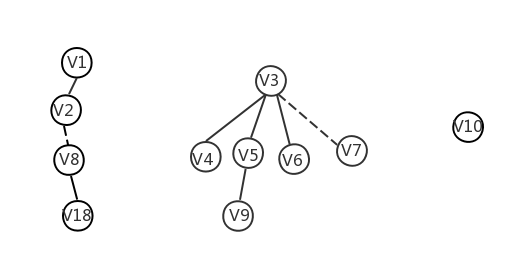}
\vspace{-0.1in}
\caption{For the graph in Fig \ref{insertmatching}, after inserting the matching $E_m$, the \emph{exPath-Tree} of $E_m$ consists of K-Path-Tree of vertex $v_1,v_3,v_{10}$.Notice that $v_1,v_2$ are in the same tree.}
\label{kpt}
\vspace{-0.2in}
\end{figure}

\subsection{Identifying Vertices with Core Number Changes after Insertion/Deletion of A Matching}
We next identify the set of vertices whose core numbers change after a matching is inserted into/deleted from the graph. Some notations will be first defined. Consider the scenario defined as follows: given a graph $G=(V,E)$ and an edge set $E_s$ = \{$e_1,e_2,...,e_s$\}, w.l.o.g., assume that for each $e_i$ = $<u_i,v_i>$, $core_G(v_i) \ge core_G(u_i) = k_i$, where $s>0$, $1\le i\le s$, and $k_i\ge 1$. After $E_s$ is inserted into or deleted from $G$, $G$ becomes a new graph $G'=(V,E')$. 

Next, we define the $k$-Path-Tree and exPath-Tree in the following Definitions~\ref{Def:KPT_u} and \ref{Def:exPath-Tree}. As shown later, after inserting/deleting a matching, only vertices on the exPath-Tree will change the core number.

\begin{definition}[\textbf{K-Path-Tree}]\label{Def:KPT_u}
For the new graph $G'=(V,E')$, $\forall u\in V$, assume $core_G(u) = k$, the \emph{$K$-Path-Tree} of $u$ is a DFS tree rooted at $u$ and each vertex $w$ in the tree satisfies $core_G(w) = core_G(u)$. For simplicity we use $T^{G'}_k(u)$ to represent the K-Path-Tree of $u$ in $G'$.
\end{definition}

\begin{definition}[\textbf{exPath-Tree}]\label{Def:exPath-Tree}
For the new graph $G'=(V,E')$ and the edge set $E_s$, the union of K-Path-Tree for every $u_i$ is called the \emph{exPath-Tree} of $E_s$, denoted as $exPT_{G'}(E_s)$.
\end{definition}

Fig.\ref{kpt} shows an example of the \emph{$K$-Path-Tree} and \emph{exPath-Tree}.

We first consider the case of inserting/deleting a $k$-matching. 

\begin{lemma}\label{Lem:kmatching}
Given a graph $G=(V,E)$, if a $k$-matching $E_k=\{e_1,e_2,...,e_p\}$ is inserted into or deleted from $G$, where $k\geq 1$, we have
 
$(i)$ only vertices in $exPT_{G'}(E_k)$ may change the core number, where $G'$ is the graph obtained from $G$ after inserting/deleting $E_k$;

$(ii)$ the change is at most 1.
\end{lemma}

\begin{IEEEproof}
We prove (i) first.
In the insertion case, we first prove that if a vertex $w\in V$ such that $core_G(w) = x \neq k$ can not change its core number. Assume $core_G(w)$ increases to $x+1$, denote the max-k-core of $w$ after insertion is $H_u^+$, before insertion is $H_u$. We know that $H_u$ is a $x-$core and $H_u^+$ is a $x+1$-core. Then at least one of the new edges must belong to $H_u^+$, as otherwise $H_u$ would be a $x+1$-core before insertion. Assume $e_i = <u_i,v_i>\in H_u^+$, and the core number change of $w$ is caused by the core number change of $u_i$. Then we have $k_i\ge x+1$ since $e_i\in H_u^+$ and $core_{G'}(v_i) \ge core_{G'}(u_i)> x+1$. Removing the k-matching $E_k$ from $G$ will decrease the degree of $u_i,v_i$ by 1 and decrease their core number to at least $x+1$. This implies $H_u^+\ E_k$ is a $x+1$-core before insertion, which is a contradiction. So if a vertex has a core number not equal to $k$, it cannot change core number.

Then we prove that if a vertex $w\in V$ such that $core_G(w) = x\in K_k$ but $w\notin exPT(E_k)_{G'}$ can not change its core number. Assume $w$ increase its core number to $x+1$. We can have that $w$ must either have a new neighbor or at least one of its neighbors increases core number. Obviously $w$ do not have any new neighbor and we have proved that vertices whose core number is not equal to $k$ won`t increase core number. So $w$ must have a neighbor whose core number is $k$ and increases core number. Applying this recursively, we will finally reach a vertex $u$ whose core number change due to gaining a new neighbor, and vertices on the recursive path all have a core number $k$, say, $w$ is in the $T_{k}(u)_{G'}$, which is a contradiction.

We have proved that for a vertex $w$, if $core_G(w)\neq k$ or $core_G(w) = k$ but $w\notin exPT(E_k)_{G'}$ can not change its core number, so only vertices in the $exPT(E_k)_{G'}$ may have their core numbers increased, the insertion proof is completed.

Then the deletion case can be proved based on the insertion. We first insert the matching and then remove them from $G'$, we can have only vertices who increase core numbers after the insertion will decrease core number back to what they are in $G$. That is to say, only vertices in the $exPT(E_k)_{G'}$ can have their core numbers decreased.

(ii) can be easily gained since $E_k\subseteq E_M$ and Lemma \ref{Lem:matchinginsert} has proved inserting/deleting a matching $E_M$ makes all vertices change core number by at most 1.
Combining all above together, the Lemma is proved.
\end{IEEEproof}


Based on Lemma \ref{Lem:kmatching}, we next consider the case of inserting/deleting a matching.

\begin{lemma}\label{Lem:exPT}
Given a graph $G=(V,E)$, if a matching $E_M = \{e_1,e_2,...,e_m\}$ is inserted into or deleted from $G$, then we can have that only vertices $w$ on $exPT_{G'}(E_M)$ can change the core number and the change is at most 1.
\end{lemma}

\begin{IEEEproof}
It is easy to see that inserting/deleting a matching $E_M$ can have the same result as inserting/deleting the k-matchings in $E_M$ one by one. Assume $E_M=E_{k_1}\cup E_{k_2}\cup, ...,\cup E_{k_p}$ and $k_i<k_j$ where $1\le i<j\le p$. We insert/delete the k-matchings one by one, and denote the graph after inserting/deleting $E_{k_i}$ as $G_{k_i}$.

First for the insertion case, when inserting $E_{k_i}$, by Lemma \ref{Lem:kmatching}, we can know that, only vertices on $exPT_{G_{k_{i}}}(E_{k_i})$ can increase the core number by at most 1. Here notice that the core numbers of some vertices on $exPT_{G_{k_{i}}}(E_{k_i})$ may be $k_i-1$ in $G_{k_{i-2}}$ and increase by 1 to $k_i$ after inserting $E_{k_{i-1}}$. We denote these vertices as $B_{k_{i-1}}$. To prove our result, we need to show that vertices in $B_{k_{i-1}}$ will not increase the core number any more. This can be obtained by Lemma \ref{Lem:matchinginsert}, since every vertex can change the core number by at most 1 after the insertion of a matching. 

The deletion case can be proved using a similar argument as above. The proof is completed.
\end{IEEEproof}

\section{Incremental Core Maintenance}\label{sec:in}
In this section, we present the parallel algorithm for incremental core maintenance after inserting an arbitrary edge set $E_I$ to graph $G=(V,E)$. Let $V_s$ denote the set of vertices connecting to edges in $E_I$. We define the \emph{maximum insertion degree} $\Delta_I$ as the maximum number of edges inserted to each vertex in $V$.

\begin{algorithm} [htb]\label{Alg:MatchingInsert}
\caption{MatchingInsert($G,E_I,V_s,core()$)}
\textbf{Input}\\
The graph, $G=(V,E)$;\\
The inserted edge set, $E_I$;\\
The set of vertices $V_I$ connected to edges in $E_I$\;
The core number $core(v)$ of each vertex in $G$;\\
\textbf{Initially} $\mathcal{C}\gets$ {empty core set}\;
\nl Properly color the inserted edges in $E_I$ using $\Delta_I+1$ colors $\{1,2,\ldots,\Delta_I+1\}$ by executing the coloring algorithm in \cite{coloring}\;
\nl $max_c\gets${the max color of edges in $E_I$}, $c\gets{1}$\;
\nl \While{$c\le max_c$}{
\nl     $E_M\gets${edges in $E_I$ with color $c$}\;
\nl     insert $E_M$ into $G$, $G$ becomes $G_c$\;
\nl     \For{each edge $e$ in $E_M$}{
            \If{core number of $e$ is not in $\mathcal{C}$}{
                add core number of $e$ to $\mathcal{C}$\;
            }
        }
\nl     \For{each core number $k$ in $\mathcal{C}$ in parallel}{
\nl         $E_k\gets${edges in $E_M$ with core number $k$}\;
            $V_c\gets${$K$-MatchingInsert($G_c$,$E_k$,$core()$)}\;
        }
\nl     \For{each vertex $v\in\cup_{k\in\mathcal{C}}V_c$} {
            $core(v) \gets{core(v)+1}$\;
        }
\nl     $c\gets{c+1}$\;
    }
\end{algorithm}

\textbf{Algorithm.} The detailed algorithm is given in Algorithm \ref{Alg:MatchingInsert}. The algorithm consists of two main steps. At first, a preprocessing procedure is executed, to split the edges into several matchings. Then the algorithm is executed in iterations in each of which a matching is inserted into the graph, and the set of vertices that change core numbers are found. 

More specifically, the preprocessing is done by properly coloring the inserted edges, such that any pair of edges with common endpoint receive different colors. It is easy to see that the edges with the same color constitute a matching. 

In the second step, as shown in Lemma~\ref{Lem:matchingdelete}, when inserting a matching into the graph, the core number of every vertex can increase by at most one. This means that we only need to find the set of vertices that increase core numbers because of the insertion of each particular edge, and the union of these vertices is just the set of vertices which will increase the core number by one. Hence, in each iteration, the edges are processed in parallel. But it deserves to pointing out that we do not make each edge processed by a processor. Instead, the edges with the same core number in the inserted matching are processed on a processor. This is to decrease duplicated processing of vertices. By Lemma~\ref{Lem:kmatching}, the edges in a $k$-matching, i.e., the edges with core number $k$ in the inserted matching, affect the core number change of the same set of vertices, those on $exPT_{G_c}(E_k)$, where $E_k$ is the $k$-matching. And by Lemma~\ref{Lem:matchingdelete}, each vertex can increase the core number by at most one. This means that when a vertex is determined to increase its core number, after inserting an edge in $E_k$, it is unnecessary to visit it any more when inserting other edges in $E_k$. Hence, by processing the edges with the same core number on a process can greatly decrease unnecessarily repeated visiting to vertices.

\begin{algorithm}  [htb]\label{Alg:insertK}
\caption{K-MatchingInsert($G_c, E_k,$ core())}
\textbf{Input}\\
The current new graph, $G_c = (V,E_c)$;\\
The inserted edges with core number $k$, $E_k$;\\
The current core number $core(v)$ of each vertex $v$\;
\textbf{Initially}, {$S\gets$ empty stack}\;
for each vertex $v\in V$, $visited[v]\gets{false}, removed[v]\gets{false}, cd[v]\gets{0}$\;
\nl compute $SD$ value for each vertex $v$ in $exPT_{G_c}(E_k)$\;
\nl \For{each $e_i=<u_i,v_i> \in E_k$}{
\nl \lIf{$core(u_i)\geq core(v_i)$}{
$r\gets{v_i}$}
\lElse{$r\gets{u_i}$}
\nl \If{$visited[r]$ = false and $removed[r]$ = false}{
\nl     \lIf{$CSD[r] =$ 0} {compute $CSD[r]$}
        \lIf{$cd[r] >=$ 0} {$cd[r]\gets{CSD[r]}$}
    \lElse
        {$cd[r]\gets{cd[r]+CSD[r]}$}
        $S.push(r)$\;
        $visited[r]\gets{true}$\;
\nl     \While{S is not empty} {
            $v\gets{S.pop()}$\;
\nl         \If{$cd[v]> k$}{
\nl             \For{each $<$ $v,w$ $>$ $\in E_c$}{
\nl                 \If{$core(w) = k$ and $SD(w) > k$ and $visited[w]$ = false}{
                       $S.push(w)$\;
                       $visited[w] \gets{ true}$\;
                       \If{$CSD[w] =$ 0} {compute $CSD[w]$}
                        $cd[w] \gets{ cd[w] + CSD[w]}$
                    }
                }
             }
\nl         \Else {
                \If{$removed[v]$=false}{
							InsertRemove($G_c,core(),cd[],\newline removed[],k,v$)
						}
            }
        }
    }
}
\nl \For{each vertex $v$ in $G$}{
    \If{$removed[v]$=false and $visited[v]$ = true}{
         $V_c \gets{V_c\cup\{v\}}$
    }
}
\nl \textbf{return} $V_k$;
\end{algorithm}
\begin{algorithm}[htb]\label{Alg:InsertRemove}
\caption{InsertRemove($G_c,core(),cd[],removed[],k,r$)}
\textbf{Input}\\
The current new graph, $G_c = (V,E_c)$;\\
The  $core(v),cd[v],removed[v]$ of each vertex\;
The core number $k$ and root $r$;\\
\nl $S\gets{ empty\ stack}$\;
\nl $S.push(r)$, $removed[r] \gets{true}$\;
\nl \While{S is not empty}{
        $v\gets{S.pop()}$\;
        \For{each $<$$v,w$$>$ $\in E_c$}{
            \If{$core(w) = k$}{
\nl             $cd[w]\gets{cd[w]-1}$\;
                \If{$cd[w] = k$ and $removed[w]$ = false}{
\nl                 $S.push(w)$\;
\nl                 $removed[w] \gets{ true}$\;
                }
            }
        }
    }
\end{algorithm}

Algorithm \ref{Alg:insertK} is used to search the vertices with core number changes when inserting a $k$-matching $E_k$. Algorithm \ref{Alg:insertK} first computes $SD$ values for vertices on $exPT_{G_c}(E_k)$, then handles the insertion of these edges one by one. In particular, for each edge $e_i$ = $<$ $u_i,v_i$ $>$ $\in E_k$, it conducts a $DFS$ search from the root which is the one in $\{u_i,v_i\}$ with smaller core number (if $core(u_i)=core(v_i)$, the tie is broken arbitrarily). Each vertex is maintained a value $cd$ which counts the number of neighbors that may help it increase the core number. The initial value of $cd$ of a vertex is set as its constraint superior degree. Vertices in $exPT_{G_c}(E_k)$ that are connected to the root are pushed into the stack, if they satisfy that the $SD$ value is larger than $k$. Only these vertices are possible to increase the core number by the definition of $SD$. Then we use the condition that $cd\leq k$ to determine that a particular vertex is impossible to increase the core number. Specifically, at every time, a vertex $v$ is fetched from the stack. If $cd[v]> k$ which means $v$ may be in a $k$+1-core, its neighbors are then visited. Otherwise, $v$ cannot be in a $k$+1-core, and a \emph{negative} DFS search as shown in Algorithm \ref{Alg:InsertRemove} are executed to remove it and spread the influence to other vertices caused by the removal of $v$, i.e., update the $cd$ values of other vertices.   

\smallskip
\textbf{Performance Analysis.}
In this part, we analyze the correctness and efficiency of our incremental algorithm. To depict the complexity of the algorithm, we first define some notations. 


For graph $G=(V,E)$, the inserted edge set $E_I$ and a subset $S$ of $E_I$, let $G_S=(V,E\cup S)$ and $K(G_S)$ be the set of core numbers of vertices in $G_S$.

For $G_S$, let $L_S = \max_{S'\subset E_I\setminus S}\max_{u\in V}\{CSD_{G_{S\cup S'}}(u)-core_{G_S}(u),0\}$. As shown later, $L_S$ is the maximum times a vertex $u$ can be visited by $InsertRemove$ procedure in the algorithm in the iteration when inserting edges to $G_S$. 

Let $K(G_S)$ denote the set of core numbers of vertices in $G_S$. For a $k\in K(G_S)$, let $V_S(k)$ be the set of vertices with core number $k$, and $N(V_S(k))$ be the neighbors of vertices in $V_S(k)$. Let $n_S=\max\{|V_S(k)|: k\in K(G_S)\}$.

Denoted by $E[V_S(k)]$ the set of edges in $G_S$ that are connected to vertices in $V_S(k)\cup N(V_S(k))$. Then we define a parameter $m_S$ as follows, which represents the maximum number of edges travelled when computing $SD$ after inserting edges to $G_S$. 
\begin{equation*}
m_S=\max_{k\in K(G_S)}\{|E[V_S(k)]|\}.
\end{equation*}

With the above notations, we prove the correctness and bound the running time of our algorithm in the following Theorem \ref{InsertCorrectness}.
\begin{theorem}\label{InsertCorrectness}
Algorithm \ref{Alg:MatchingInsert} can update the core numbers of vertices in $O(|E_I|*|V_I|+\Delta_I* \max_{S\subseteq E_I}\{m_S + L_S*n_S\})$ time, after inserting an edge set $E_I$ to a graph $G$, where $V_I$ is the set of vertices in $G$ connecting to edges in $E_I$ and $\Delta_I$ is the maximum insertion degree.
\end{theorem}
\begin{IEEEproof}
We first prove the correctness. As discussed before, a proper coloring can split the inserted edges into multiple matchings. When inserting a matching into the graph, it is only necessary to find the vertices that change their core numbers and increase their core numbers by one. Hence, the edges in a matching can be processed in parallel. This constitutes the base of our parallel algorithm. Then when processing a particular matching $E_k$ in a process,  by Lemma~\ref{Lem:kmatching} and Lemma~\ref{Lem:exPT}, only vertices in $exPT_{G_c}(E_k)$ are possible to increase the core number. Hence, Algorithm~\ref{Alg:insertK} visit all vertices that may change the core number. Furthermore, based on the definitions of $CSD$ and $cd$ and Lemma~\ref{le:inserne}, if $cd(v)<k$ for a vertex $v$, it is impossible for $v$ to increase the core number. Therefore, the algorithm removes vertices that will not increase the core number definitely. And the negative DFS process in Algorithm \ref{Alg:InsertRemove} ensures to spread the influence of a removed vertex, i.e., update the $cd$ values (which count the number of neighbors that may support a particular vertex's core number change) of vertices in $exPT_{G_c}(E_k)$. By Lemma \ref{Alg:MatchingInsert}, each vertex being determined to increase the core number is unnecessary to visit any more, as shown in Algorithm \ref{Alg:insertK}, since these vertices will not increase their core numbers any more. Finally, after executing the algorithm, vertices that are visited but not removed will increase their core numbers, as these vertices all have at least $k+1$ neighbors and they constitute $k+1$-cores. Combining the above together, the correctness of the algorithm is guaranteed.

As for the time complexity, there are two stages in the algorithm: first, the inserted edges are split into matchings by coloring, and second, in each iteration, edges in a matching with the same core number are inserted to the graph and the core numbers of vertices are updated. As shown in  \cite{coloring}, the coloring of inserted edges takes $O(|E_I|*|V_I|)$ time. We next analyze the time used in the second stage.

We first bound the number of iterations. In each iteration, the edges with one particular color are processed. As shown in \cite{coloring}, the number of colors used is at most $\Delta_I+1$. Hence, the number of iterations is also upper bounded by $\Delta_I+1$.

Now consider an iteration $i$ in the second stage of the algorithm execution. Denote by $G_i$ the graph obtained after iteration $i-1$ as $G_i$, and by $E_M$ the matching gained in iteration $i$. The computation of $CSD$ values for vertices in $exPT_{G_c}(E_M)$ takes $O(m_{E_M})$ time. In the algorithm, each vertex is visited for once to determine whether to update its core number. But it needs to notice that each vertex may be visited for multiple times in the negative DFS procedures that disseminate the influence of a removed vertex. However, if a vertex $v$ is visited in the negative DFS procedure, $cd(v)$ is decreased by 1. Hence, each vertex can be visited by at most $L_S$ times, since a vertex will be removed if its $cd$ value is decreased to its core number. Then we can get that the total time for an iteration is upper bounded by $O(m_{E_M}+L_{E_M}*n_{E_M})$.

By all above, the running time of the whole algorithm can be bounded as stated.
\end{IEEEproof}

\section{Decremental Core Maintenance}\label{sec:de}
The algorithm for decremental core maintenance, as given in Algorithm~\ref{Alg:MatchingDelete}, is very similar to the incremental one. The only difference is that we use Lemma~\ref{corollary:sd} instead of Lemma~\ref{le:inserne} to determine whether a vertex will decrease the core number after deleting edges.  The maximum number of edges deleted from each vertex is defined as the \emph{maximum deletion degree}, denoted as $\Delta_D$.

\begin{algorithm}  [htb]\label{Alg:MatchingDelete}
\caption{MatchingDelete($G, E_D, V_s, core()$)}
\textbf{Input}\\
The graph, $G=(V,E)$;\\
The deleted edge set, $E_D$;\\
The set of vertices $V_D$ connected to edges in $E_D$\;
The core number $core(v)$ of each vertex in $G$;\\
\textbf{Initially} $\mathcal{C}\gets$ {empty core set}\;
\nl Properly color the deleted edges in $E_D$ using $\Delta_D+1$ colors $\{1,2\ldots,\Delta_D+1\}$ by executing the coloring algorithm in \cite{coloring}\;
\nl $max_c\gets${the max color of edges}, $c\gets{1}$\;
\nl \While{$c < max_c$}{
\nl     $E_M\gets${edges in $E_D$ with color $c$}\;
\nl     delete $E_M$ from $G$, $G$ becomes $G_c$\;
\nl     \For{each edge $e$ in $E_M$}{
            \If{core number of $e$ is not in $\mathcal{C}$}{
                add core number of $e$ to $\mathcal{C}$\;
            }
        }
\nl     \For{each core number $k$ in $\mathcal{C}$ in parallel}{
\nl         $E_k\gets${edges in $E_M$ with core number $k$}\;
            $V_c\gets${$K$-MatchingDelete($G_c,E_k,core()$)}\;
        }
\nl     \For{each vertex $v$ in $\cup_{k\in\mathcal{C}}V_c$} {
            $core(v) \gets{core(v)-1}$\;
        }
\nl     $c\gets{c+1}$\;
    }
\end{algorithm}

\begin{algorithm}  [htb]\label{Alg:deleteK}
\caption{$K$-MatchingDelete($G_c, E_k, core()$)}
\textbf{Input}\\
The current new graph, $G_c = (V,E_c)$;\\
The edges with core number $k$, $E_k$;\\
The core number $core(v)$ of each vertex in $V$\;
\textbf{Initially}, {$S\gets$ empty stack}\;
for each vertex $v\in V$, $visited[v]\gets{false}, removed[v]\gets{false}, cd[v]\gets{0}$\;
\nl \For{each $e_i=<u,v> \in E_k$}{
\nl \lIf{$core(u)\geq core(v)$}{
$r\gets{v}$}
\lElse{$r\gets{u}$}
\nl \If{$core(v) \neq core(u)$}{
        \If{$visited[r]$ = false}{
            $visited[r] \gets true$\;
            $cd[r] \gets SD(r)$\;
        }
\nl         \If{$removed[r]$ = false}{
            \If{$cd[r] < k$}{
\nl                DeleteRemove($G_c,core(),cd[],removed[],k,r$)
            }
        }
    }
\nl    \Else{
\nl        \If{$visited[u]$ = false}{
                $visited[u] \gets {true}$\;
                $cd[u] \gets{ SD[u]}$\;
            }
\nl        \If{$removed[u]$ = false}{
\nl            \If{$cd[u] < k$}{
                 DeleteRemove($G_c,core(),cd[],removed[],k,u$)
                }
            }
\nl         \If{$visited[v]$ = false}{
                $visited[v] \gets{true}$\;
                $cd[v] \gets{SD(v)}$\;
            }
\nl         \If{$removed[v]$ = false}{
                \If{$cd[v] < k$}{
\nl                DeleteRemove($G_c,core(),cd[],removed[],k,v$)
                }
            }
        }
    }
\nl \For{each vertex $v$ in $G$}{
    \If{$removed[v]$ = true and $visited[v]$ = true}{
         $V_c \gets{ V_c\cup\{v\}}$ }
    }

\nl \textbf{return} $V_c$;
\end{algorithm}

\begin{algorithm}[htb]\label{Alg:DeleteRemove}
\caption{DeleteRemove($G_c,core(),cd[],removed[],k,r$)}
\textbf{Input}\\
The current new graph, $G_c = (V,E_c)$;\\
The  $core(v),cd[v],removed[v]$ of each vertex\;
The core number $k$ and root $r$;\\
\nl {$S\gets{empty\ stack}$}\;
\nl $S.push(r)$, $removed[r] \gets{true}$\;
\nl \While{S is not empty}{
        $v\gets{S.pop()}$\;
        \For{each $<$ $v,w$ $>$ $\in E_c$}{
\nl         \If{$core(w) = k$}{
\nl             \If{$visited(w)$ = false}{
                    $visited[w] \gets{true}$\;
                    $cd[w] \gets{cd[w] + SD(w)}$\;
                }
\nl             $cd[w]\gets{cd[w]-1}$\;
                \If{$cd[w]$ $<$ $k$ and $removed[w]$ = false}{
\nl                $S.push(w)$\;
\nl                 $removed[w] \gets {true}$\;
                }
            }
        }
    }
\end{algorithm}

\textbf{Performance Analysis.}
The correctness and efficiency of the proposed decremental algorithm can be analyzed similarly as the incremental one.
At first, we define some notations. 

For graph $G=(V,E)$, the deleted edge set $E_D$ and a subset $R$ of $E_D$, let $G_R=(V,E\setminus R)$ and $K(G_R)$ be the set of core numbers of vertices in $G_R$. 

For $G_R$, let $F_R = \max_{R\subset E_D\setminus R}\max_{u\in V}\{SD_{G_{R\cup R'}}(u)-core_{G_R}(u)+1,0\}$.

For $k\in K(G_R)$, let $V_R(k)$ be the set of vertices with core number $k$ and $n_R=\max\{|V_R(k)|: k\in K(G_R)\}$.

Denote by $E(V_R(k))$ the set of edges connected to vertices in $V_R(k)$. We then define $m_R$ as follows, 
\begin{equation*}
m_R=\max_{k\in K(G_R)}\{|E(V_R(k))|\}.
\end{equation*}
$F_R$, $n_R$ and $m_R$ will depict the time used in each iteration in the algorithm execution. 

Using a similar argument as that for analyzing the incremental algorithm, we can get the following result, which states the correctness and efficiency of the decremental algorithm. 

\begin{theorem}\label{DeleteCorrectness}
Algorithm \ref{Alg:MatchingDelete} can update the core numbers of vertices in $O(|E_D|*|V_D|+|\Delta_D*\max_{R\in E_D}\{m_R+F_R*n_R\})$ time, after deleting an edge set $E_D$ from a graph $G$, where $V_D$ is the set of vertices in $G$ connecting to edges in $E_D$ and $\Delta_D$ is the maximum deletion degree.
\end{theorem}

\section{Experiment Studies}\label{sec:experiment}
In this section, we evaluate the performances of our algorithms by experiments. The experiments use three synthetic data sets and seven real-world graphs, as shown in Table \ref{table_graph}.

There are two main variations in our experiments, the original graph and the inserted/deleted edge set. We first evaluate the efficiency of our algorithms on real-world graphs, by changing the size and core number distribution of inserted/deleted edges. Then we evaluate the scalability of our algorithms using synthetic graphs, by keeping the inserted/deleted edge set the same and changing the size of synthetic graphs. Besides, we show the parallelization of our algorithms for several typical graphs, the number of iterations and the number of edges handled per iteration are showed. At last, we compare our algorithms with the state-of-the-art sequential core maintenance algorithms, TRAVERSAL algorithms given in \cite{Sar2016Incremental}, to evaluate the speedup ratio of our parallel algorithms. The comparison experiments are conducted on four typical real-world data sets.

All experiments are conducted on a Linux machine having 8 Intel Xeon E5-2670@2.60GHz CPUs with support for 16 concurrent threads and 64 GB main memory. The algorithm is implemented in C++ and compiled with g++ compiler using the -O3 optimization option.

\smallskip
\textbf{Data sets.} We use seven real-world graphs and random graphs generated by three different models. The seven real-world graphs can be downloaded from SNAP \cite{SNAP}, including social network graphs (LiveJournal, Youtube, Gowalla), collaboration network graphs (DBLP, ca-condmat), communication network graphs (WikiTalk) and Web graphs (web-BerkStan). The synthetic graphs are generated by the SNAP system using three models: the Erd\"os-R\textbf{$\acute{e}$}nyi (ER) graph model \cite{ER}, which generates a random graph; the Barabasi-Albert (BA) preferential attachment model \cite{BA}, in which each vertex has $k$ connected edges; and the R-MAT (RM) graph model \cite{RM}, which can generate large-scale realistic graphs similar to social networks. For all generated graphs, the average degree is fixed to 8, such that when the number of vertices in the generated graphs is the same, the number of edges is the same as well.

\begin{figure}
 \subfigure[Real-world Graphs]{
    \label{core1} 
    \includegraphics[width=1.6in]{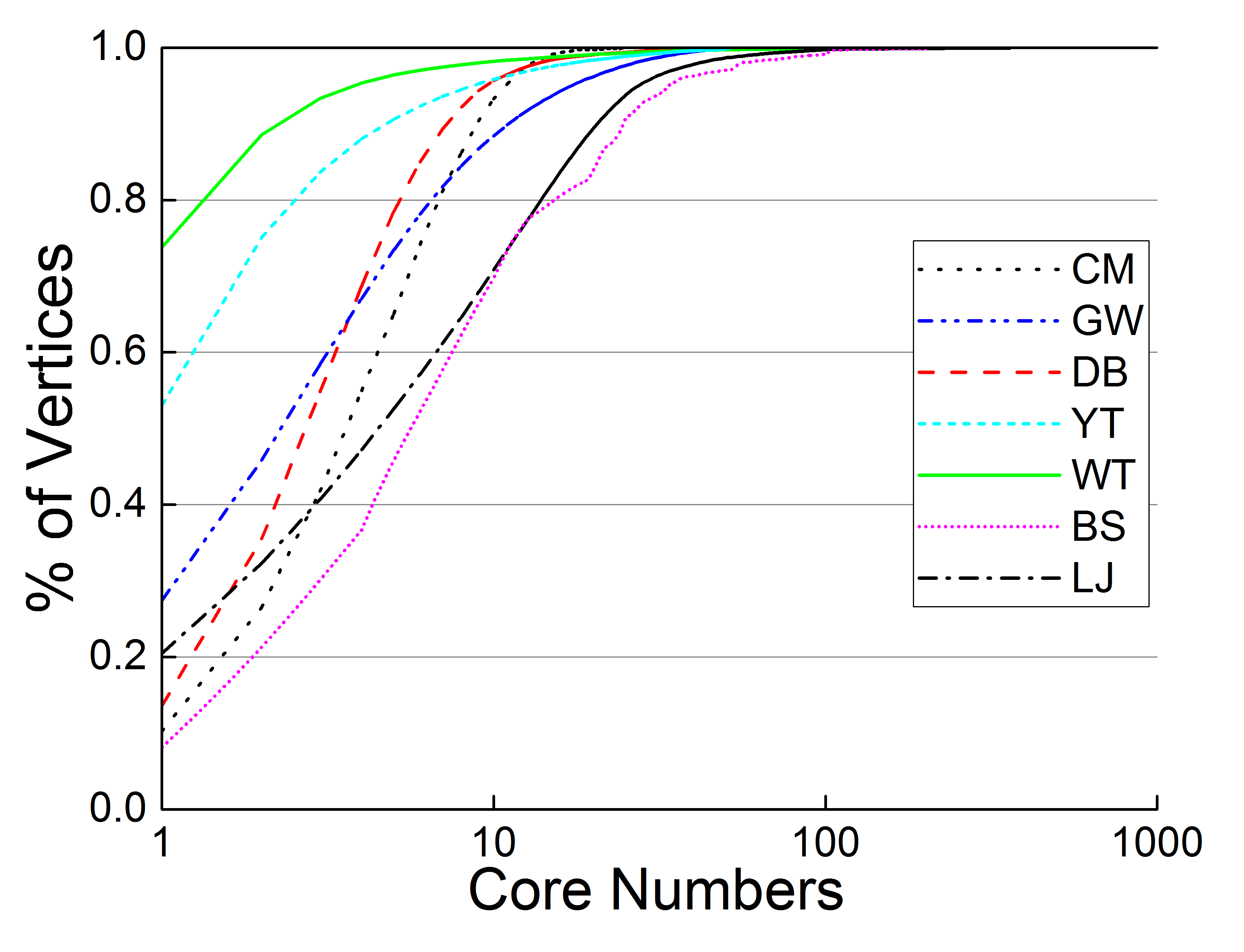}
    }
 \subfigure[Generated Graphs]{
    \label{core2} 
    \includegraphics[width=1.6in]{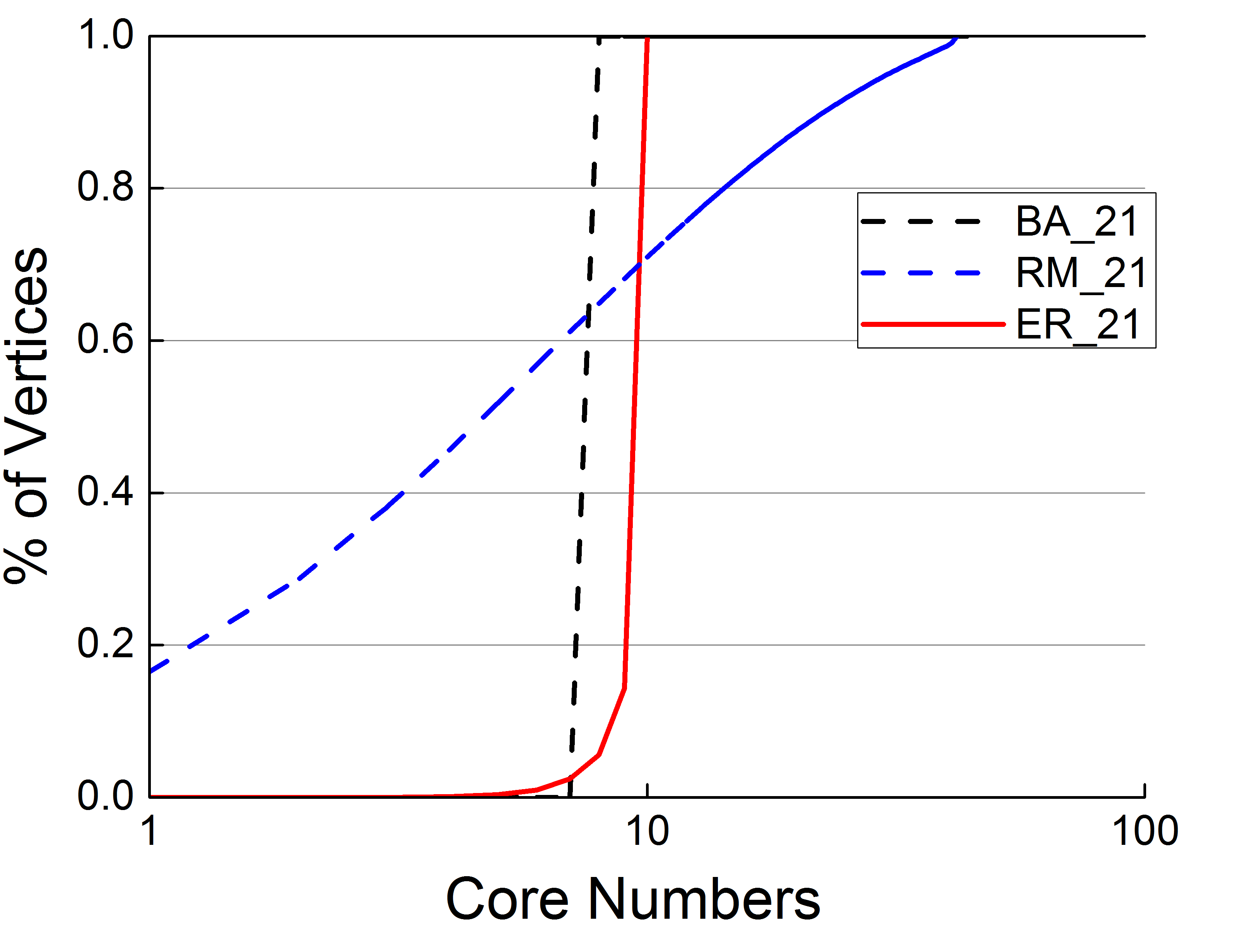}
    }
 \caption{Core Distribution}
 \label{cores} 
\end{figure}

Fig. \ref{core1} and Fig. \ref{core2} show the core number distributions of the seven real-world graphs and the generated graphs with $2^{21}$ vertices. From Fig. \ref{core1}, it can be seen that in real-world graphs, most of the vertices have core numbers smaller than 10. Especially, in WT, more than 70\% of vertices have core number 1. For the generated graphs, as shown in Fig. \ref{core2}, all vertices have a core number of 8 in the BA graphs. In the ER graph, the max core number of vertices is 10, and almost all vertices have core numbers close to the max one. In the RM graph most vertices have small core numbers which are more close to real-world graphs, and as the core number $k$ increases the number of vertices with core number $k$ decreases. As shown later, the core number distribution of a graph will affect the performances of our algorithms.

We use the \emph{average processing time per edge} as the efficiency measurement of the algorithms, such that the efficiency of the algorithms can be compared in different cases.

\begin{table}
\renewcommand{\arraystretch}{1.3}
\caption{Real-world graph datasets}
\centering
\begin{tabular}{|c|c|c|c|}
\hline
Datasets & n=$|V|$ & m=$|E|$  & max core\\
\hline
CM(ca-condmat)  & 23.1K     & 93.5K & 25  \\
GW(Gowalla)     & 196.6K     & 1.9M & 51  \\
DB(DBLP)        & 0.31M     & 1.01M & 113 \\
YT(YouTube)     & 1.13M     & 1.59M & 35    \\
WT(wiki-Talk)   & 2.4M      & 9.3M  & 131   \\
BS(web-BerkStan)& 0.68M     & 13.3M & 201   \\
LJ(LiveJournal) & 4.0M      & 34.7M & 360   \\
\hline
\end{tabular}
\label{table_graph}
\end{table}

\subsection{Performance Evaluation}
We evaluate the impacts of three factors on the algorithm performance: the number of inserted/deleted edges, the core numbers of edges inserted/deleted, and the size of the original graph. The first factor mainly influences the iterations needed to process the inserted/deleted edges, and the other two factors affect the processing time in each iteration. We use the real-world graphs to conduct the first two evaluations, and the synthetic graphs on the third one.

\begin{figure}
 \subfigure[Insertion]{
    \label{Size_ins} 
    \includegraphics[width=1.6in]{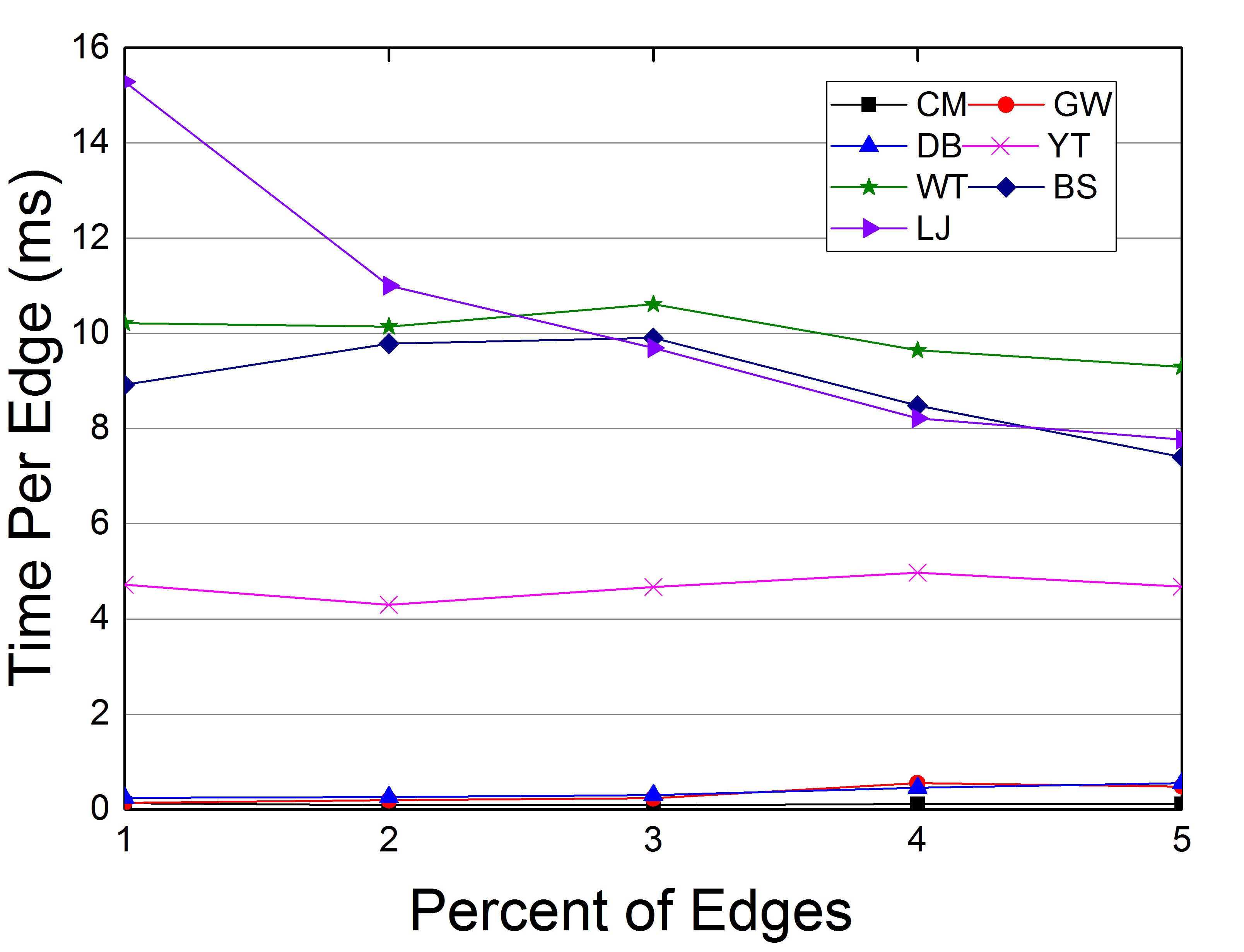}
    }
 \subfigure[Deletion]{
    \label{Size_del} 
    \includegraphics[width=1.6in]{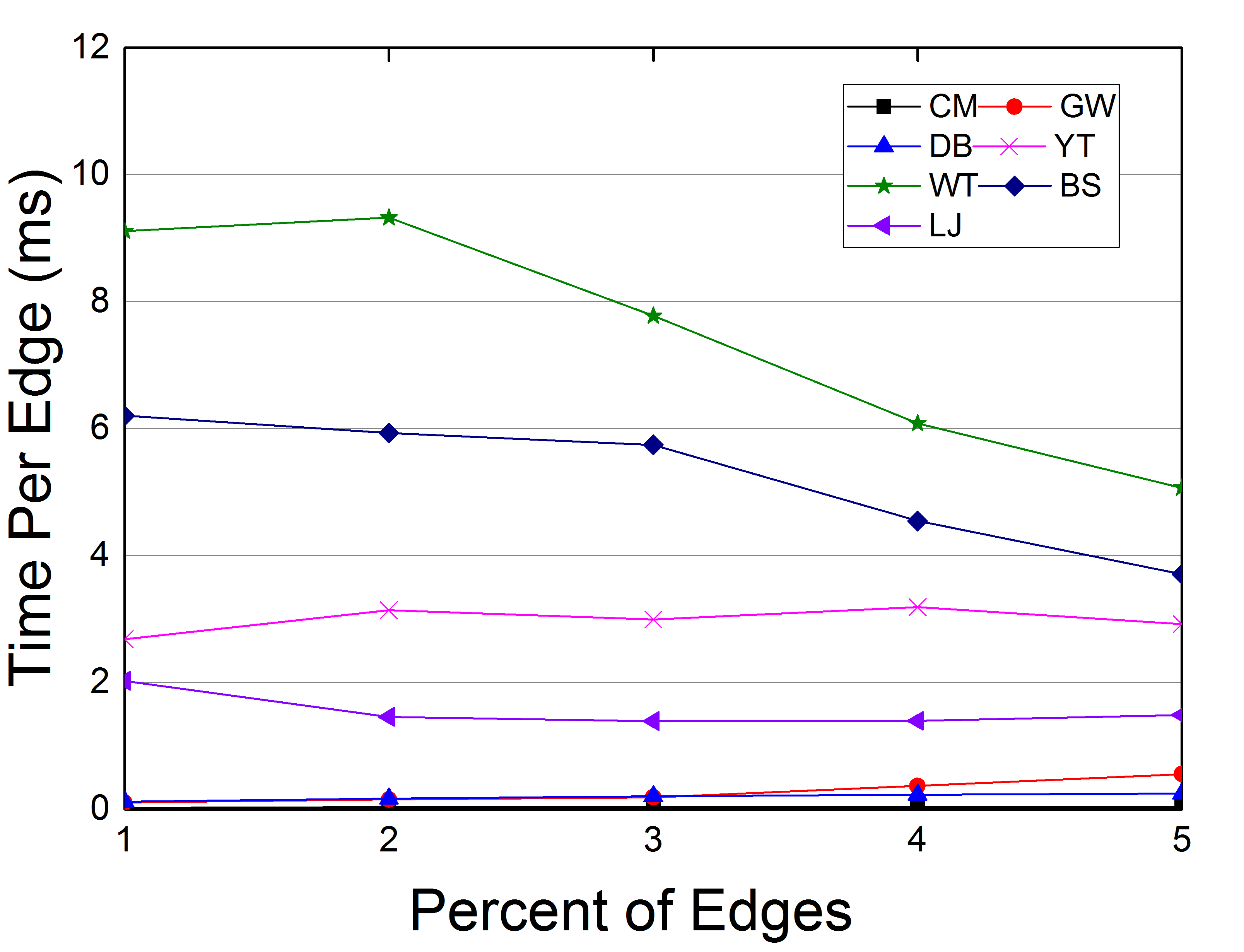}
    }
 \caption{Impact of Inserted/Deleted Edge Number}
 \label{SizeChange} 
 \end{figure}

First we change the number of updated edges. We randomly insert/delete $P_i$ edges from the original graph, where $P_i=i\%$ for $i=1,2,3,4,5$. Fig. \ref{Size_ins} and Fig. \ref{Size_del} show the processing time per edge for insertion and deletion cases respectively. It can be seen that the processing time per edge is less than $10ms$ in the deletion case and less than $16ms$ in the insertion case for all graphs. Besides it is found that, when more edges are updated, the time needed per edge is decreased in general. This is because our algorithms have better parallelism in the case of large amount of updates. In this case, more edges can be selected into the matching processed in each iteration. Therefore, our algorithms are more suitable for processing large amount of updates.

\begin{figure}
 \subfigure[Insertion]{
    \label{core_ins} 
    \includegraphics[width=1.6in]{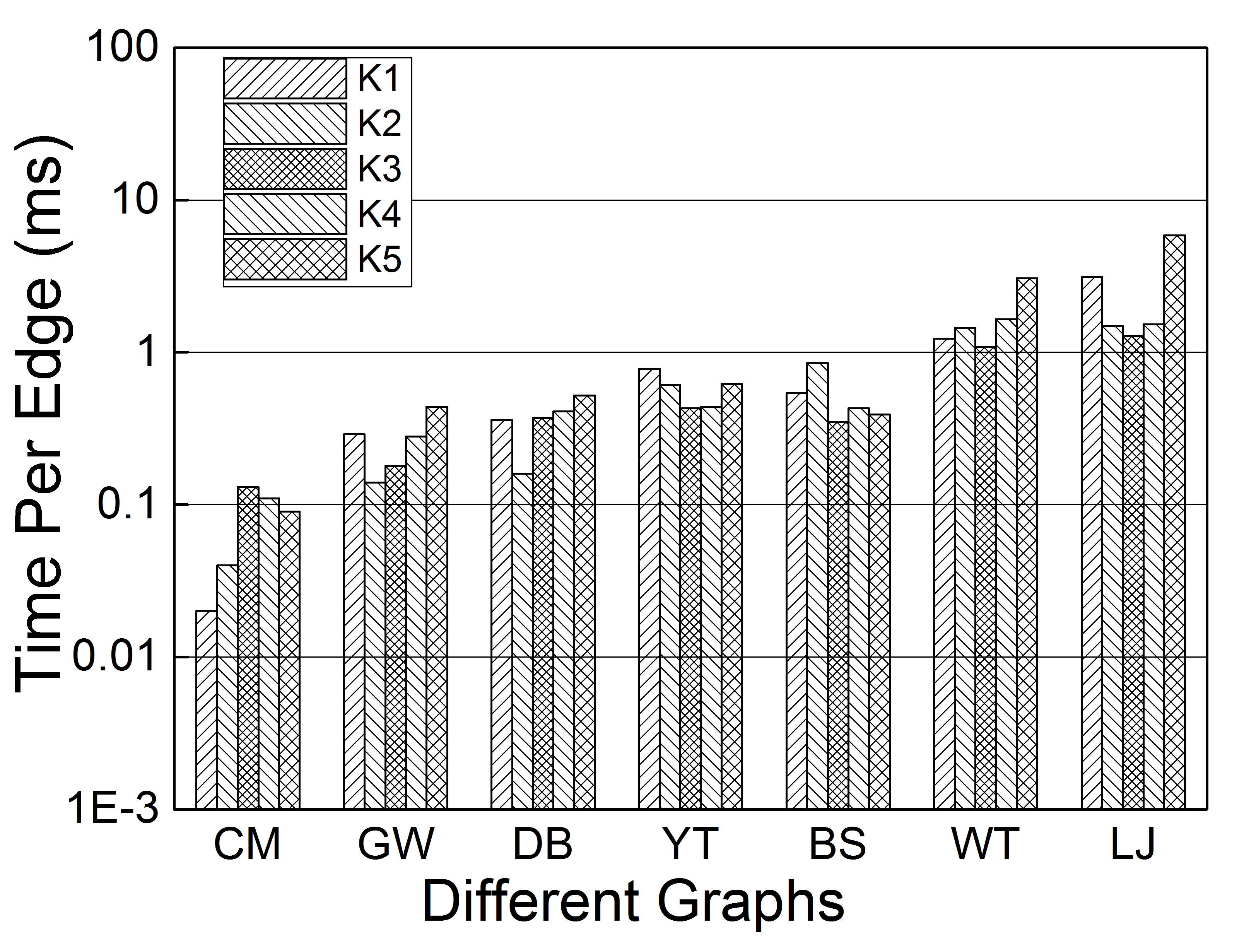}
    }
 \subfigure[Deletion]{
    \label{core_del} 
    \includegraphics[width=1.6in]{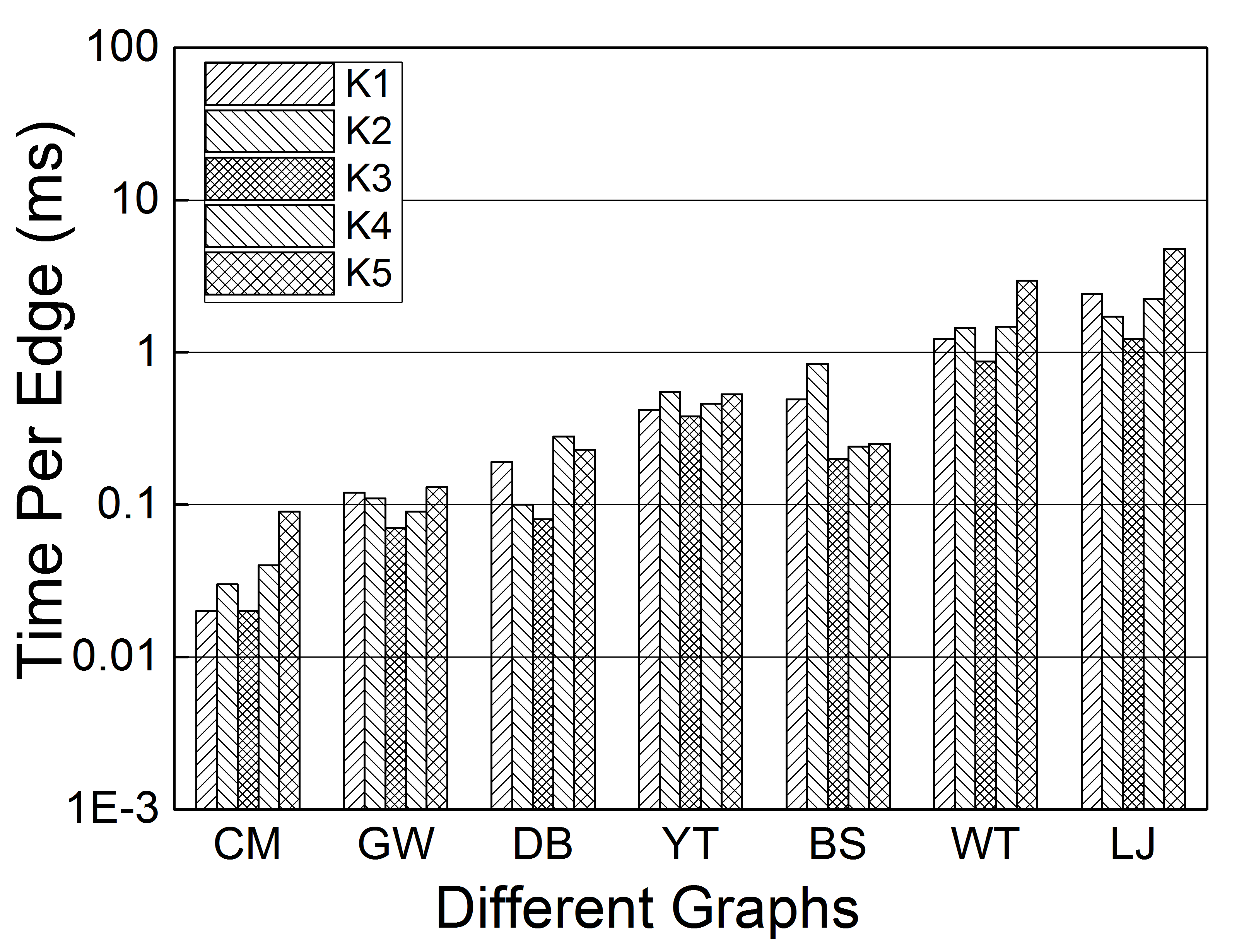}
    }
 \caption{Impact of Core Number of Inserted/Deleted Edges}
 \label{coreChange} 
\end{figure}

Then, we vary the core number distributions of the update edges to evaluate the performance of our algorithms. This kind of tests can help us understand how our algorithms behave over changes on different areas of the graph. The results are illustrated in Fig. \ref{coreChange}. By the core distributions showed in Fig. \ref{core1}, we choose five typical core numbers \{$K1,K2,K3,K4,K5$\} in an increasing order for each of the seven graphs. For each core number, we randomly select 20\% edges of that core as the update edge set. It can be seen from Fig. \ref{coreChange} that larger core number needs a larger average processing time for most of the graphs. This is because though the amount of vertices with larger core number is small, each vertex has a large degree, so these vertices tend to have more neighbors when we choose the edges randomly. While in our algorithm a vertex may decrease its degree by only one in an iteration, so more iterations will be needed, which results in longer running time. However the running time per edge does not vary so much when the core number of edge changes, which demonstrates that our algorithms are stable upon changes on different parts of the graph.

\begin{figure}
 \subfigure[Insertion]{
    \label{graph_ins} 
    \includegraphics[width=1.6in]{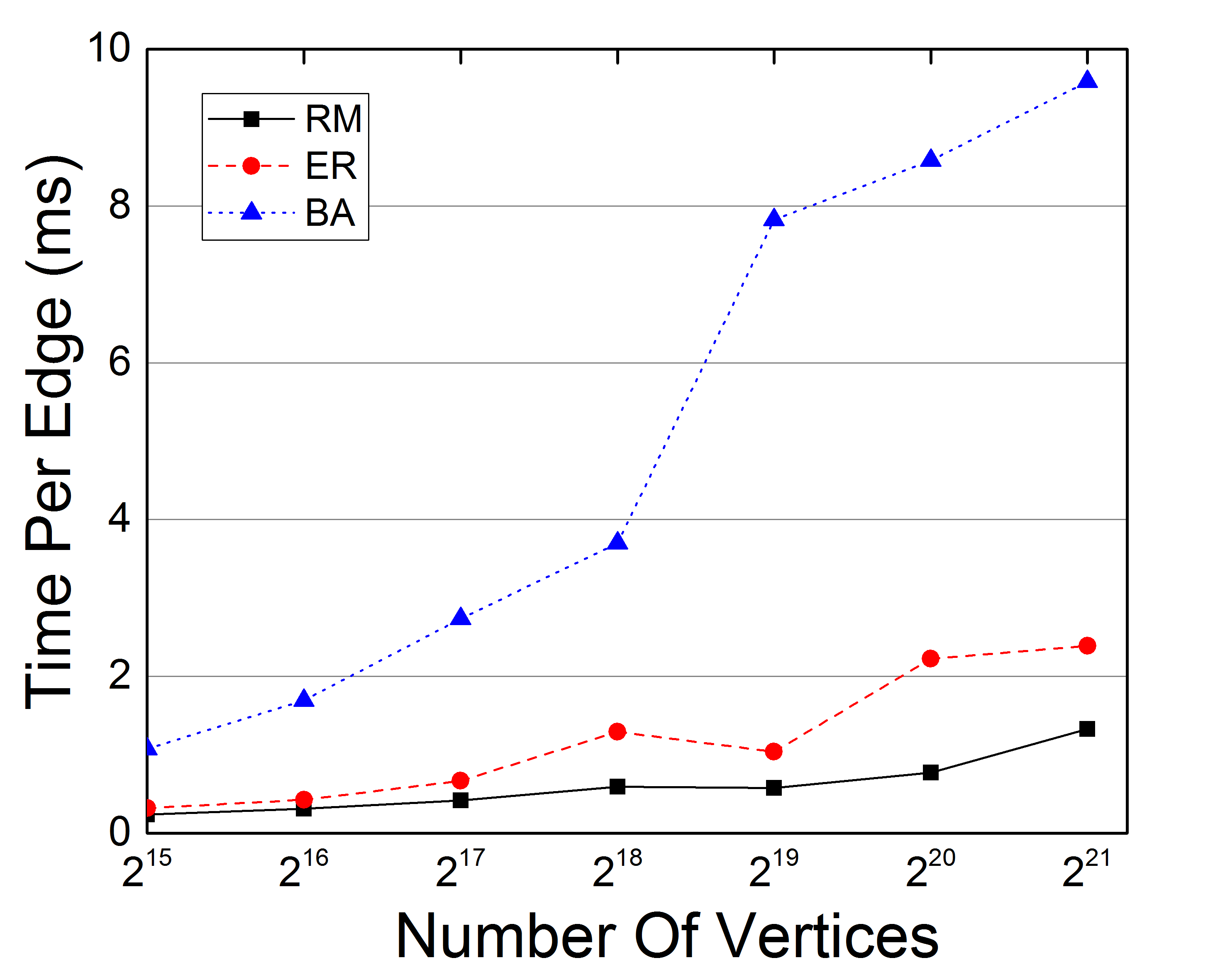}
    }
 \subfigure[Deletion]{
    \label{graph_del} 
    \includegraphics[width=1.6in]{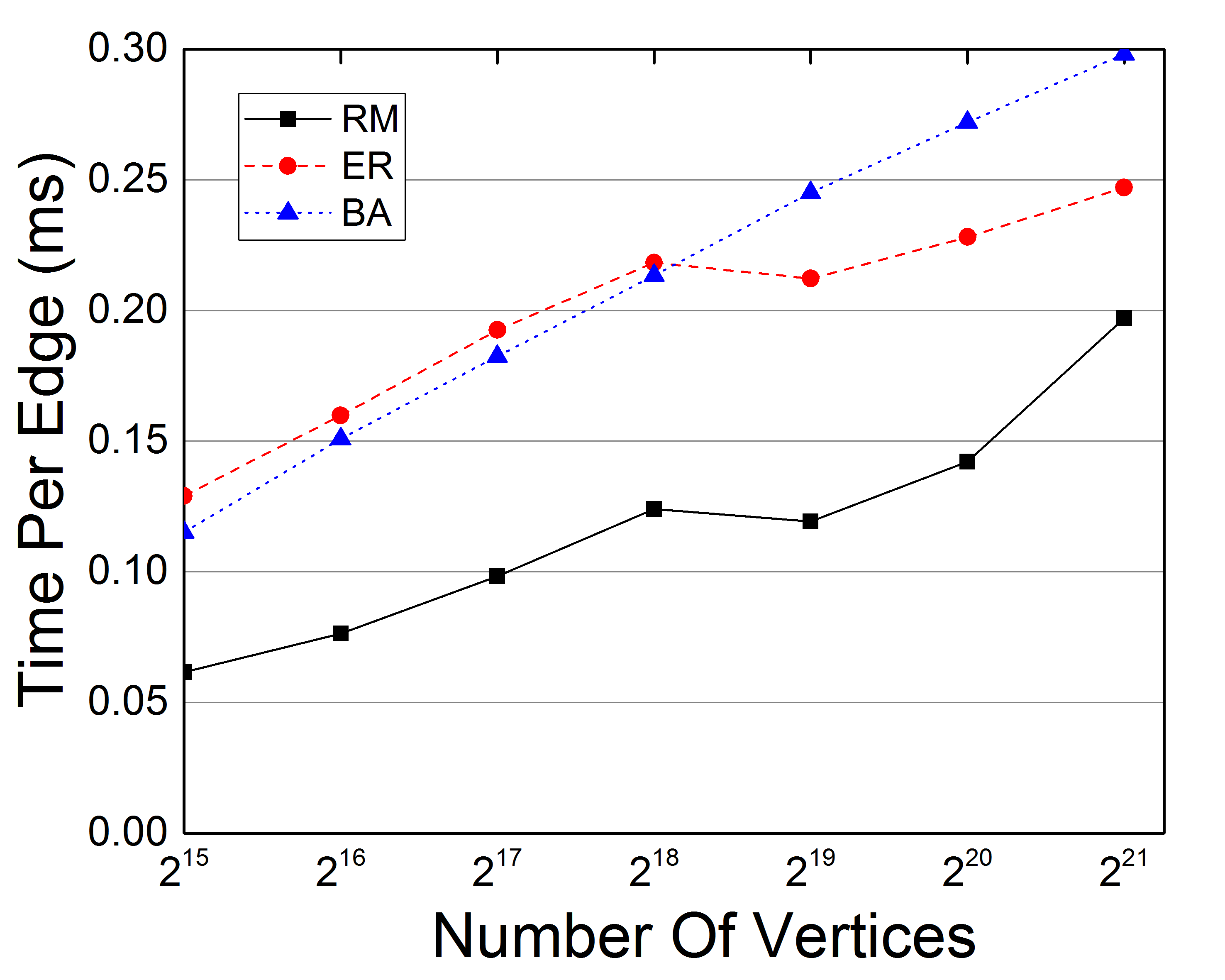}
    }
 \caption{Impact of Orginal Graph Size}
 \label{graphChange} 
 \end{figure}

\begin{figure}
 \subfigure[Number of Edges Processed in Each Iteration]{
    \label{edgenum} 
    \includegraphics[width=1.6in]{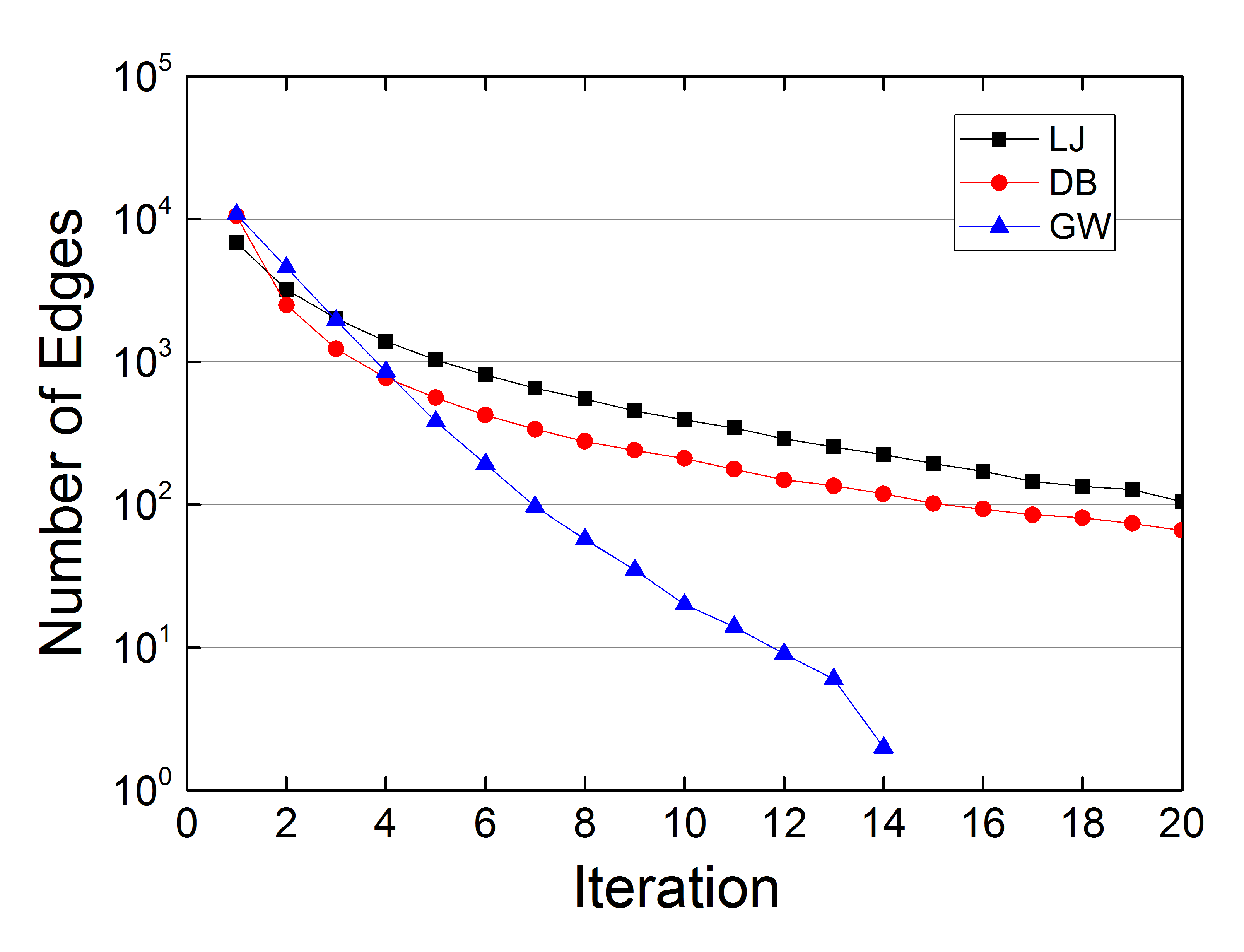}
    }
 \subfigure[Number of Threads in Each Iteration]{
    \label{thread} 
    \includegraphics[width=1.6in]{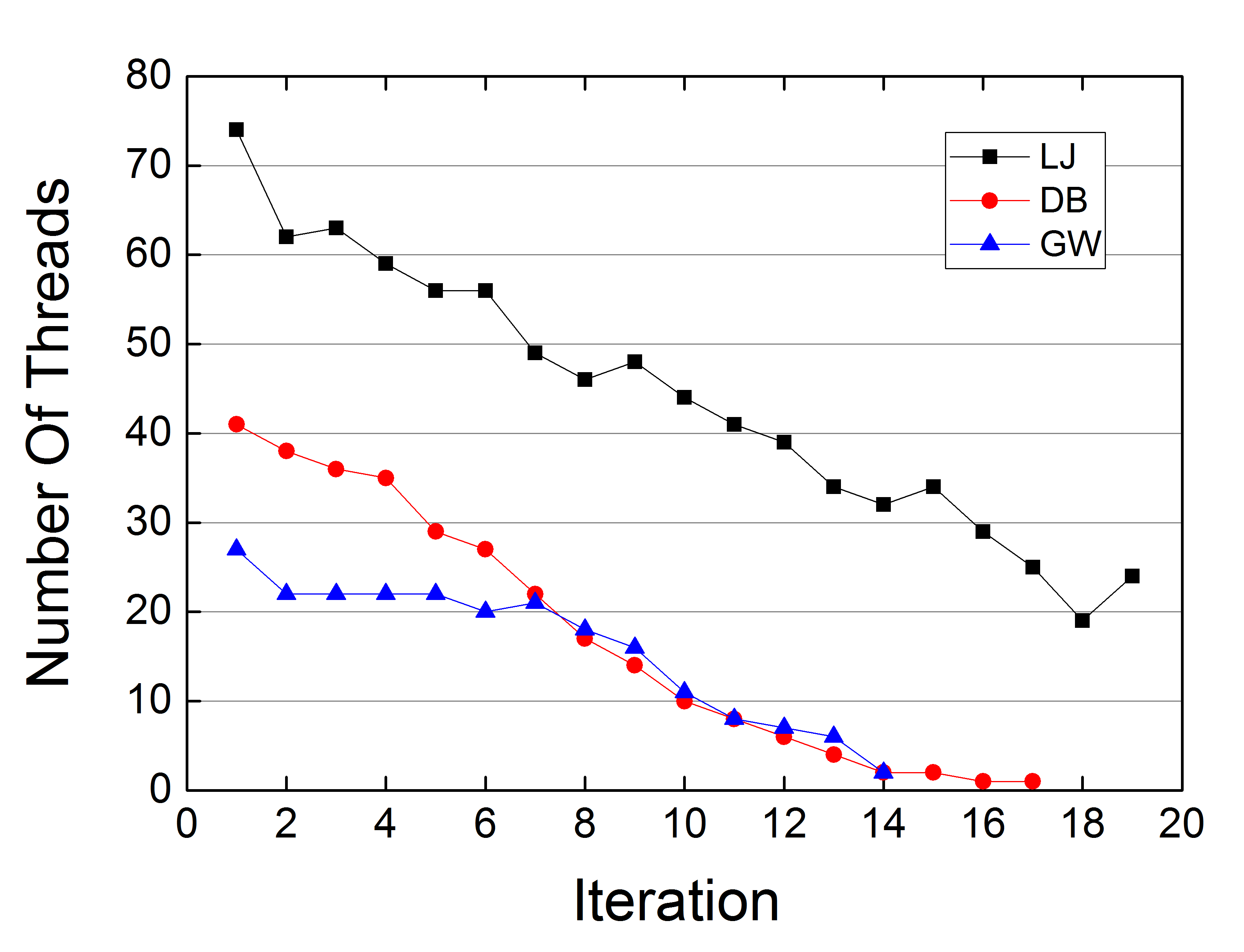}
    }
 \caption{Parallelism Exhibition}
 \label{parallel} 
\end{figure}

\begin{figure}
 \subfigure[Insertion]{
    \label{cmp_ins} 
    \includegraphics[width=1.6in]{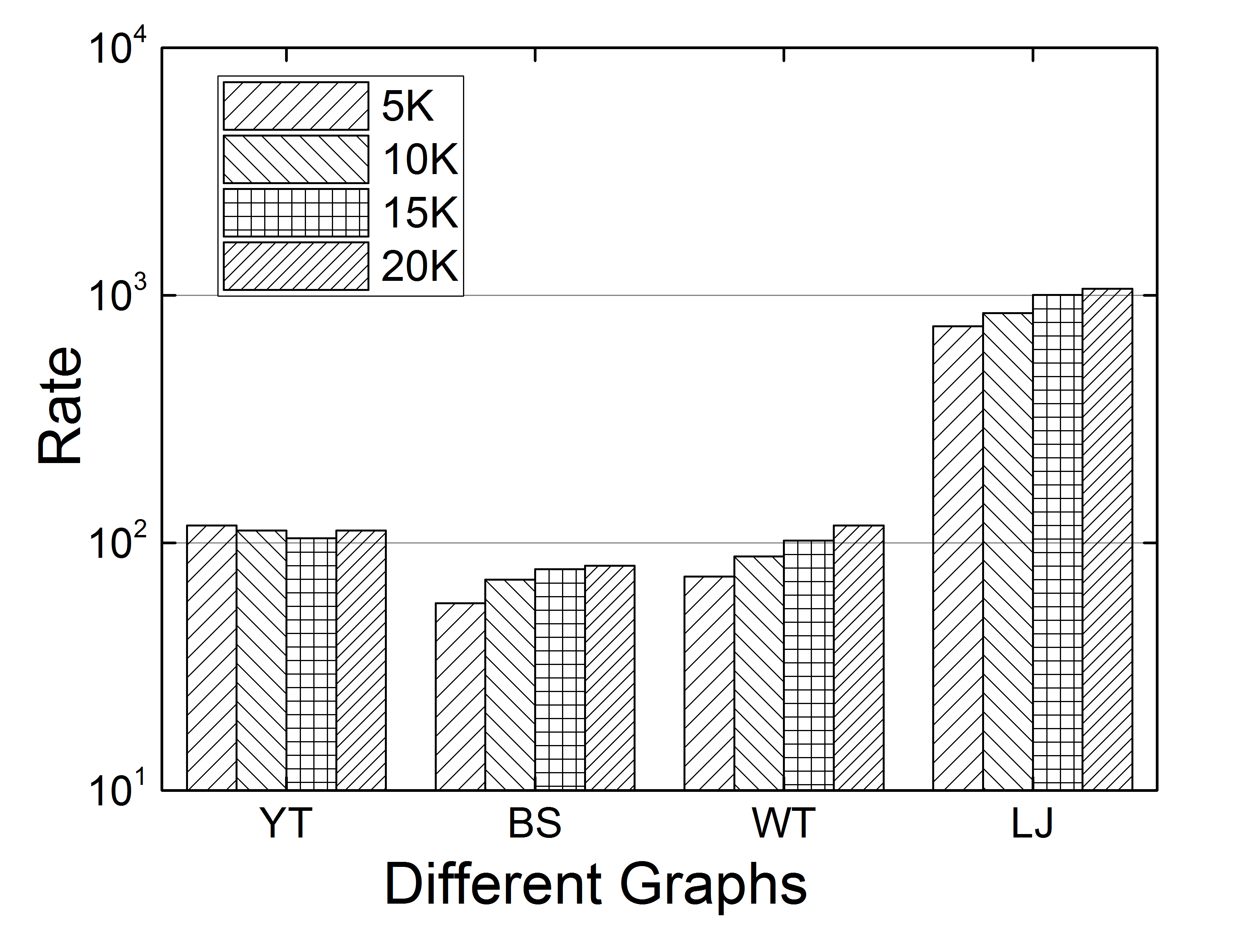}
    }
 \subfigure[Deletion]{
    \label{cmp_del} 
    \includegraphics[width=1.6in]{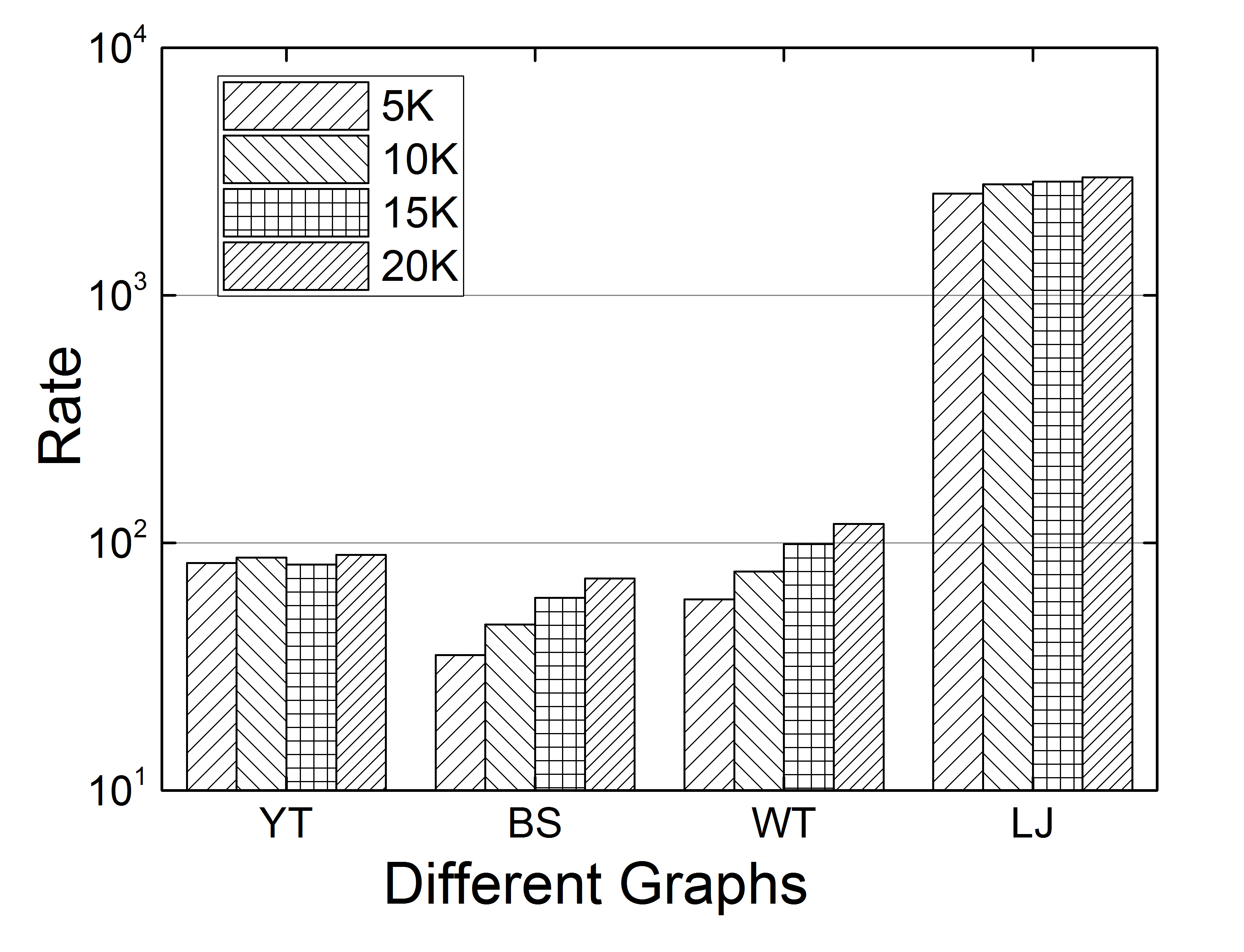}
    }
 \caption{Comparison with the TRAVERSAL Algorithm}
 \label{Comparison} 
 \vspace{-0.1in}
\end{figure}

Additionally, we test the scalability of our algorithms using the three synthetic datasets. With the average degree fixed as 8, we vary the number of vertices $|V|$ from $2^{15}$ to $2^{21}$. For each graph, we randomly select 10000 edges from each graph and show the processing time per edge as the graph size ranges. It can be seen from  Fig. \ref{graphChange} that when the graph size increases at an exponential rate, the processing time just increases linearly. This indicates that our algorithms is stable upon different graph size and can handle graphs with extremely large size in acceptable time. As the figure shows, the processing time for BA graph is a bit longer than the other two graphs. This is because in BA graphs all vertices have the same core number 8, the parallelism of our algorithms is very poor in this case, since initially there is only one processor used in the algorithm execution. This situation can also be seen as the worst case for our algorithms. However, real-world graphs do not exhibit this property as shown in Fig. \ref{core1} and \ref{core2}.

At last, we exhibit the parallelism of our algorithms on the three graphs LJ, DB and GW repectively, which is shown in Fig \ref{parallel}. Here, we randomly choose 20000 edges as the update edges. In Fig \ref{edgenum}, the x-axis represents the iteration number during the algorithm execution and the y-axis represents the number of edges that are processed in one iteration. It can be seen that for all three graphs, almost 10000 edges are selected into the matching in the first iteration and can be processed. Besides, for graphs DB and GW, more than 18000 edges are handled in the first three iterations and there are less than 20 iterations in total. For the graph LJ, due to the edge distribution over vertices, the insertion/deletion degree is larger, and hence there are more iterations. However, we can see that more than 1000 edges are processed in each of the first 6 iterations. As the iteration increases, it can be found that the number of edges processed in each iteration decreases. This is because less vertices still have unprocessed edges as the algorithm execution, and only one edge connected to every vertex is selected into the matching in each iteration. Hence, the number of selected edge decreases as the algorithm execution. The number of threads used in each iteration is shown in Fig \ref{thread}.

\subsection{Comparisons with Sequential Algorithms}
We next evaluate the speedup ratio of our parallel algorithms, comparing with algorithms that sequentially handle edge insertions/deletions. We compare with the state-of-the-art sequential algorithms, \textbf{TRAVERSAL} algorithms given in \cite{Sar2016Incremental}. The comparison is conducted on four typical real-world graphs, YT, BS, WT and LJ as given in Table \ref{table_graph}. For each graph, we randomly select \{5K, 10K, 15K, 20K\} edges as the update set. The comparison results are illustrated in Fig. \ref{cmp_ins} and Fig. \ref{cmp_del}. In the figures, the y-axis represents the speedup ratio of our algorithm over the \textbf{TRAVERSAL} algorithms.

Fig. \ref{cmp_ins} and Fig. \ref{cmp_del} show that in almost all cases, our algorithms achieve a speedup ratio of around $100$ in both incremental and decremental core maintenance. Especially for the large-scase graph LJ, the speedup ratio is more than 1000. It can also be found that the speedup ratio increase as the number of edges inserted/deleted and the graph size increase. Hence, our algorithms provide better efficiency in large-scale graphs and processing large amount of graph changes.

\textbf{Evaluation Summary.} The experiment results show that our algorithms exhibit good stability and scalability. Comparing with sequential algorithms, our algorithms speed up the core maintenance procedure significantly. Additionally, our algorithms are suitable for handling large amount of edge insertions/deletions in large-scale graphs, which is desirable in realistic implementations.

\section{Conclusion}\label{sec:conclusion}
In this paper, we present the parallel algorithms for core maintenance in dynamic algorithms. Our algorithms exhibit significant acceleration comparing with sequential processing algorithms that handle inserted/deleted edges sequentially. Experiments on real-world and synthetic graphs illustrate that our algorithms implement well in reality, especially in scenarios of large-scale graphs and large amounts of edge insertions/deletions.

\end{document}